# Spin Processor

A. K. Khitrin[*] and V. L. Ermakov[†]

*Department of Chemistry and Biochemistry, University of Oklahoma,
Norman, Oklahoma 73019-0370, USA;*
[†]on leave from *Kazan Physical-Technical Institute, Kazan 420029, Russia.*

**It is experimentally demonstrated that with multi-frequency excitation of weak amplitude the dynamics of a cluster of dipolar-coupled spins can be manipulated to perform parallel logic operations with long bit arrays.**

Quantum computing is realized by controlled dynamic evolution at times, shorter than the quantum decoherence time of a chosen system. The most general experimental scheme can be viewed as excitation-detection, where the excitation carries information about the input data and algorithm, and the detection produces an output information. Quantum bits are naturally represented by spins ½ and, therefore, quantum information processing can be thought of as spin dynamics, when the result of quantum computation is a response of a system of coupled spins to some complicated time-dependent excitation.

Let us consider a cluster of $N$ coupled spins ½ with the Hamiltonian of internal interactions $\mathcal{H}$. For simplicity, we suppose that all of its $2^N$ eigenvalues $\varepsilon_i$ are not degenerate and that there are no coinciding differences $\omega_{ij} = \varepsilon_i - \varepsilon_j$. The Hamiltonian of interaction with the excitation field $\Omega(t)$, applied at a single frequency $\omega_0$ in x-direction is (in frequency units) $\Omega S_x \cos(\omega_0 t)$, where $S_x$ is the projection of the total spin of the system on x-axis. The maximum number of different frequencies in the response signal is the number of frequencies $\omega_{ij}$ corresponding to non-zero matrix elements $(S_x)_{ij}$. For a spin cluster with dipolar couplings, this number $C_{2N}^{N+1} \sim 2^{2N}$ is comparable to the total number of pairs of states.

---

[*]e-mail: khitrin@ou.edu

Depending on the amplitude of $\Omega$, different types of response signal can be expected.

1. At $\Omega \to 0$, or when $\Omega$ is much less than a difference $\Delta\omega$ between neighbor frequencies $\omega_{ij}$, a response signal at a single frequency $\omega_0 = \omega_{ij}$ can be excited. This excitation can be described by a dynamics of the corresponding two-level system. Very roughly, $\Delta\omega$ can be estimated as $\Delta\omega \sim 2^{-2N} \omega_{loc}$, where $\omega_{loc}$ is the total width of the spectrum.

2. At $\Delta\omega \ll \Omega \ll 2^N \Delta\omega$, many transitions are excited simultaneously, but most of them have no common quantum levels, so the system behaves as a sum of non-interacting two-level systems. The response in this case is that of a system with inhomogeneously broadened spectrum.

3. At $\Omega \sim 2^N \Delta\omega$ there are "islands" of coupled (through common quantum states) transitions. We have found experimentally [1, 2] that in this case a very sharp coherent response signal of large amplitude can be generated. The dynamics of such signal is quite different from what one could expect for systems with inhomogeneous broadening, and the phase of the signal is opposite to that of the normal absorption signal. This is a collective response: during a lifetime of this signal individual spins can perform thousands of almost random rotations in their fluctuating local fields.

4. At $2^N \Delta\omega \ll \Omega \ll \omega_{loc}$, a "percolation" takes place, and all quantum levels of the system are connected by excited transitions. In this case, a thermal quasi-equilibrium is established. Populations of quantum levels are described by spin temperature(s) and time evolution of the system under the influence of the excitation field is described by the thermodynamic theory of Provotorov [3].

5. Finally, at $\Omega \gg \omega_{loc}$, it is an excitation with a "hard" pulse. When the pulse is short, the response is a normal linear response signal.

For a comparatively large cluster, say $N \sim 10$, the density of allowed transitions is very high and spin-lattice relaxation prevents from observing the response signals of the types 1 and 2. Excitation of coherent long-living response signal of the type 3 at a single frequency affects only a small subspace of the total Hilbert space. This makes it possible, with multi-frequency excitation, to store and process in parallel large amount of information. If the first pulse excited some narrow region of the spectrum, the second pulse, by exciting the same region, creates non-linear interference between the two excitations, which can be used for implementing logic operations. Different narrow regions of the spectrum can be used as parallel devices.

For efficient information processing a cluster should be large enough to have complicated dynamics but small enough to remain quantum. The physical system we used is the nematic liquid crystal 4-*n*-pentyl-4'-cyanobiphenyl (5CB). Fast molecular motions average interactions between molecules, but 19 proton spins of one molecule are coupled with residual dipole-dipole interactions. Therefore, the system is a good model of an ensemble of non-interacting spin clusters.

The example of parallel logic operation with 64-bit array is shown in Fig. 1. It is an implementation of a parallel bitwise NOT operation, which flips the bits by changing ones to zeros and vice versa. Spectrum (c) in Fig. 1 represents the result of a parallel bitwise NOT operation simultaneously applied to each bit of spectrum (b). It was obtained when a pulse similar to that used for (b) was applied in anti-phase after a pulse which was used to generate spectrum (a), so that the corresponding peaks in (a) are "erased".

Finally, we would like to discuss the advantages offered by the dynamics of composite quantum systems for information processing. To make the switches in a processor work faster, one needs to decrease the number of elementary objects participating in a computation and to increase the "driving force" or external field. As we have demonstrated here, with the use of collective quantum dynamics, the number of elementary objects per one-bit operation can be substantially less than one, and the information is processed much faster than the flipping of individual spins under the influence of the applied RF field. Another advantage of this approach is the reduction of power consumption. Since the external fields are time-dependent, higher fields mean higher power. However, in practice, there is always a limit on the power which can be applied to the system without damaging it. The ability to perform parallel operations substantially decreases the required power. For example, if the bitwise operation in Fig. 1 were performed by serial flips of a single spin, it would require an average RF power two orders of magnitude higher for the same processing speed. Systems of nuclear spins are very slow but the principles considered here can be extended to other physical systems to construct much faster quantum processors.

**Acknowledgments.** The authors thank Bing M. Fung for helpful discussions.

Fig. 1. A parallel bitwise NOT gate operation using a spin cluster (5CB). Spectrum (a) is the absolute-value spectrum generated by a 50 ms pulse which is the sum of 64 harmonics. It represents the number $2^{64} - 1$ in binary notation. Spectrum (b) is the spectral representation of an arbitrarily chosen integer number $0 \leq x < 2^{64}$ obtained by applying a different 50 ms 64-frequency pulse, with some of the amplitudes set to zero as indicated below the spectrum. In decimal notation, x = 7348754808244345529. Spectrum (c) was obtained by the pulse in (a) followed by a 10 ms pulse similar to that used for spectrum (b), applied in anti-phase. It represents the number $y = 2^{64} - 1 - x$, or y = 11097989265465206086 in decimal notation. The amplitude of the RF field per harmonic was 10.5 Hz for 50 ms pulses and 2.9 Hz for the 10 ms pulse. The number of transients was 1024.

a)

11111111111111111111111111111111111111111111111111111111

b)

0110010111111000000010110010000000011101101100100011010101111001

c)

10011010000000111111010011011111110001001001101110010101000110

Frequency, Hz